\documentclass[3p,times]{jpconf}
\usepackage{graphicx} 
\begin{document}

\title{Hadron production from $\mu-Deuteron$ scattering at $\sqrt{s}=17~GeV$ at COMPASS}

\author{Astrid Morreale on behalf of the COMPASS collaboration\\National Science Foundation\\\small4201 Wilson Boulevard, Arlington, Virginia 22230, USA\\
 French Alternative Energies and Atomic Energy Commission\\ \small CEA Saclay, IRFU/SPhN, Gif sur Yvette, 91191, France}

\begin{abstract}

Hadrons proceeding from quasi-real photo-production are one of the many probes accesible at the Common Muon Proton Apparatus for Structure and Spectroscopy (COMPASS) at CERN. These hadrons provide information on the scattering between photon and partons through $\gamma$-gluon($g$) direct channels as well as $q$-$g$ resolved processes.
Comparisons of unpolarized differential cross section measurements to next-to-leading order (NLO) pQCD calculations are essential to develop our understanding of proton-proton and lepton-nucleon scattering at varying center of mass energies.  These measurements are important to asses the applicability of NLO pQCD in interpreting polarized processes.  
In this talk we will present the unidentified charged separated hadron cross-sections  measured by the COMPASS experiment at center of mass energy of $\sqrt{s}=17~GeV$,  low $Q^{2}$ (Q$^{2}<0.1~GeV^{2}/c^{2}$) and high transverse momenta (p$_{T}>1.0\,GeV/c$.)  

\end{abstract}

\section{Introduction}
\label{introduction}
Cross-sections of final state inclusive and semi-inclusive particles are benchmark measurements which may elucidate information on polarized and unpolarized photon-parton and parton-parton interactions.  
While perturbative quantum chromo dynamics (pQCD) is a well established theory of strong interactions, it does not provide an exact analytical calculation. In a pQCD factorized framework, a hadron cross-section can be defined as the convolution of three main ingredients (Eq.~\ref{formula_xsect1}) from which only one is fully calculable:  
\small
\begin{itemize}
\item The parton distribution functions $f_{a}(x_{a},\mu_F)$,  $f_{\gamma}(x_{\gamma},\mu_F)$ and $\Delta f_{a}(x_{a},\mu_F)$ at a factorization scale $\mu_F$ and momentum fraction $x_{a}$ carried from parton $a$ (or $\gamma$.)
\item The fragmentation functions (FF) $D_{h^{\pm}}(z_{b},\mu_F)$ where $z_{b}$ is the relative energy of hadron $h$ proceeding from parton $b$.\item The hard scattering cross-section of partons $d\hat{\sigma^b_{\gamma a}}(d\Delta\hat{\sigma^b_{\gamma a}})$ which is the calculable part in pQCD.  
\end{itemize}
\normalsize
To obtain information on polarized parton distributions such as that of the gluon ($\Delta G$), one could rely on measurements of asymmetries of final state particles in bins of measured transverse momentum p$_{T}$. Asymmetries are the ratio of the polarized to unpolarized cross-sections (Eq.~\ref{formula_all}). These measurements offer an elegant way of accessing parton information by counting observed particle yields in different helicity states of the interacting beam(s) and target ($++, --$\, versus\, $+-, -+$.) These ratios are normalized by the polarization in the beam and the target ($P_{B}, P_{T}$) times the dilution factor $F$. 

\scriptsize
\newcommand{\sz}{\hspace*{-10pt}}
\begin{eqnarray}\label{formula_xsect1}
d\sigma\sz\\
&\sz=\sz&\sum_{a, b= {\rm q}(\bar{\rm q}), {\rm g}}\,
\int dx_\gamma
\int dx_a
\int dz_b  \,\,
 f_\gamma (x_\gamma,\mu_F) \, f_a (x_a,\mu_F)
D_b^{h}(z_b,\mu_F') \times d\hat{\sigma^b_{\gamma a}}  \nonumber 
\end{eqnarray}

\begin{eqnarray}\label{formula_all}
 A_{LL} &\sz=\sz& 
 \frac{\displaystyle\sum_{a, b={\rm q}(\bar{\rm q}), {\rm g}} 
\Delta f_{\gamma }\otimes \Delta f_{a} \otimes \Delta \hat{\sigma}\otimes D_{h/b}} 
{\displaystyle\sum_{a, b={\rm q}, \bar{\rm q}, {\rm g}} f_{\gamma} 
\otimes f_{a} \otimes \hat{\sigma}
\otimes D_{h/b}} \\\nonumber  &=& \frac{\sigma_{++} -\sigma_{+-}}{\sigma_{++}+\sigma_{+-}} \,\, 
=\,\, \frac{1}{FP_{B}P_{T}} \frac{N^{++}-N^{+-}}{N^{++}+N^{+-}} 
\qquad
\end{eqnarray}
\normalsize

The unpolarized differential cross-section defines the denominator of the asymmetries definition in Eq.~\ref{formula_all}. These unpolarized measurements are of interest as they can add new results at different $\sqrt{s}$ and help verify applicable theoretical calculations. Cross-sections in addition can help constrain the FF's~\cite{dss} which introduce an uncertainty into these models. The aim of the current work is to contribute with measurements of semi-inclusive deep inelastic scattering (SIDIS) cross-sections at $\sqrt{s}=\,17 \,GeV$. Unpolarized SIDIS probes the number density of partons  with a fraction $x$ of the momentum of the parent nucleon. The detection of the outgoing lepton in a SIDIS measurement allows the access, among other kinematic variables, to the $Q^{2}$ and $y$ of the events.
 
\section{Hadron cross-sections at $\sqrt{s}=17\,GeV$}
\begin{figure}\centering

\includegraphics[width=15pc]{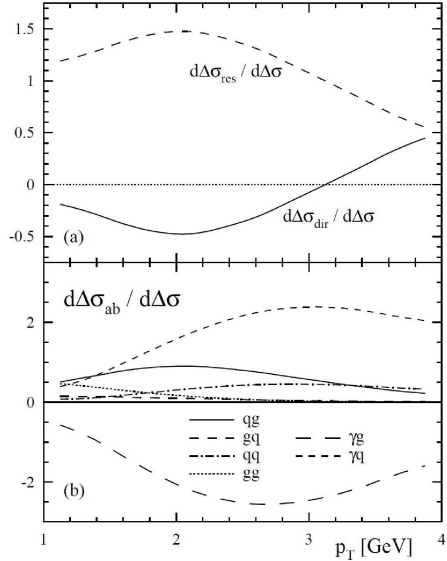}

\caption{\label{werner}Partonic contributions to quasi-real photo production. See ~\cite{jager1} and~\cite{jager2}}

\end{figure}

The reaction of interest for the current measurement is $\mu^{+} d\longrightarrow \mu^{+'}h^{\pm}X$. Final state hadrons are detected at low $Q^{2}$ (below $0.1~GeV^{2}/c^{2}$ ) and high $p_{T}$ (above 1~GeV/c). This latter ensures a large momentum transfer in the reaction~\cite{jager1}. 
Production of these final state hadrons proceed from direct photon-parton $\gamma_{dir}-(q,\bar q,g)$ and resolved photon-parton $\gamma_{res}-(q,\bar q, g)$  initiated sub-processes in $\mu-{\rm deuteron}$ center of mass (cm) collisions (Fig.~\ref{werner}.) The  $\gamma_{dir}-(q,\bar q,g)$ NLO contributions are those for which the hadrons are produced immediately after the virtual photon interacts directly with the target partons. The  $\gamma_{res}-(q,\bar q, g)$ processes are those where the photon fluctuates into partons before interacting with the target. These latter processes are akin to hadroproduction  from proton+proton (p+p) collisions such as those at RHIC or at the LHC, where the partons from the incoming beam interact with the target(beam) partons. Measurements of hadron production proceeding from p+p data have been compared to pQCD calculations at different $\sqrt{s}$ ranges ~\cite{soffer}. Results of these comparisons have shown that the pQCD calculations have not correctly described hadron production at decreasing $\sqrt{s}$. Only high  $\sqrt{s}$ seemed to be properly described by these models. The discrepancy between experiment and theory at low $\sqrt{s}$ has opened the question of pQCD applicability at these energy regimes. The aim of the present work is to measure charge hadron production with COMPASS' kinematic capabilities and compare the resulting measurements to appropiate NLO calculations. The partonic contributions leading to these hadron cross-sections at COMPASS energies have been previously calculated in~\cite{jager1} and~\cite{jager2}.

\section{The COMPASS detector at CERN}
The COMPASS experiment is a fixed target experiment at CERN which uses secondary and tertiary beams from the Super Proton Synchrotron (SPS). The muon beam used for the scattering comes from the decay of  $\pi^{+}$ which proceed from the 
scattering of a primary proton beam on a Beryllium target. The proton beam is extracted from the CERN's SPS at a cycle of approximately 17 seconds. The resulting muon beam from each cycle is then collided on a two cell $^6LiD$ target. The two stage forward spectrometer tracks particles from these interactions with a rate capability of a few MHz/channel~\cite{compass} and a spacial precision better than $100~\mu m$. Particles can be detected at very small angles ($\sim1~$mrad) and large angles ($\sim$ 120~mrad for the current data set under study.)  These scattering angles correspond to a primarily positive pseudo-rapidity coverage ($\eta_{cm}$) that can reach up to the forward values of $\sim2.4$. For the low $Q^{2}$ measurements of interest the ladder and inner triggers~\cite{compass} are used.

\section{Measuring cross-sections at COMPASS}\label{kinematics}
To measure production cross-sections,  the number of all final state charged hadrons in the data-set under interest are normalized by detector acceptances and efficiencies as well as the total integrated luminosity $\int Ldt$. The experimental definition which can be derived from Eq.~\ref{formula_xsect1} is as follows:

\scriptsize
\begin{eqnarray}
E\frac{d^{3}\sigma}{d\vec{p\,\,}^{3}} \sz
&=& \frac{d^{3}\sigma}{p_{T}d\eta dp_{T}d\phi} = \frac{1}{2\pi p_{T}} \frac{d^{2}\sigma}{dp_{T}d\eta}\,\\\nonumber&=& \frac{1}{2\pi p_{T}} \frac{N_{h^{\pm}}(p_{T})}{\int Ldt\,\epsilon_{Acc}\Delta p_{T} \Delta \eta}\label{formula_xsect2}
\end{eqnarray}\vspace{-10pt}
\normalsize

Where:
\small
\begin{itemize}\label{itemsxsec}
\item \textit{$\vec{p\,\,}^3$} is the 3 momentum of the particle.
\item $\eta$ is the pseudo-rapidity, described as $\eta =-ln[tan(\frac{\theta}{2})]$, where $\theta$ is the angle between the
 particle momentum $\vec{p}$ and the virtual photon axis. $\eta$  can be translated to the center of mass frame to:\\$\eta_{cm}=-ln(tan(asin(p_{T}/P)/2))-0.5ln(2P_{beam}/M_{proton})$.
 
\item p$_{T}$ is the transverse momentum of the particle, defined with respect to the virtual photon.
\item $\Delta$p$_{T}$ is the bin width.
\item $\phi$ is the azimuthal angle ($2\pi$ in the case of COMPASS.)
\item N$_{h^{\pm}}$ is the total number of reconstructed final state hadrons (charge separated) in a p$_{T}$ bin.
\item $\epsilon_{Acc}$ accounts for the geometrical acceptance, momentum smearing, as well as the efficiency of the reconstruction algorithm. 
\item $\int Ldt$ is the integrated luminosity. 
\end{itemize}
\normalsize
\subsection{Simulation}
A full simulation using the Monte-Carlo generator PYTHIA 6~\cite{pythia} was employed. The simulation generated $\mu^{+}$+proton and $\mu^{+}$+neutron interactions at appropriate center of mass energies. A GEANT3~\cite{geant} based program which incorporates COMPASS detector materials and interactions was used. The GEANT based program simulated the response of the detector setup for the events generated by PYTHIA. The validity of the Monte-Carlo's description of the data was evaluated by estimating the data over Monte-Carlo ratio in different kinematic and laboratory variables pertinent to the measurement. Extraction of acceptances in both 1D as a function of $p_{T}$ and 2D  were performed and results were found compatible.
\subsection{Integrated luminosity}
The integrated luminosity is obtained from the beam flux measurement using only periods with stable
data taking and good spectrometer performance. This measurement was determined~\cite{Lumi} using beam scalers~\cite{compass} on a spill by spill basis, where each spill corresponds to the beam delivery by the SPS. The integrated luminosity  was found to be : $\int Ldt=$142.4 $pb^{-1}\,\pm 10\%$. The error is systematic and it reflects all uncertainties in the measurement including changes in beam intensity which affected the beam rate calculations.  
\subsection{Radiative corrections}
Radiative corrections for the production of hadrons at the kinematic limits described in section~\ref{kinematics} were evaluated~\cite{haprad} by Andrei Afanasev~\cite{andrei} from Jefferson Lab. His first estimates show that the data sample has a radiative effect contribution of less than $\sim 2\%.$

\section{Results}
\begin{figure}
\centering
\includegraphics[width=16pc]{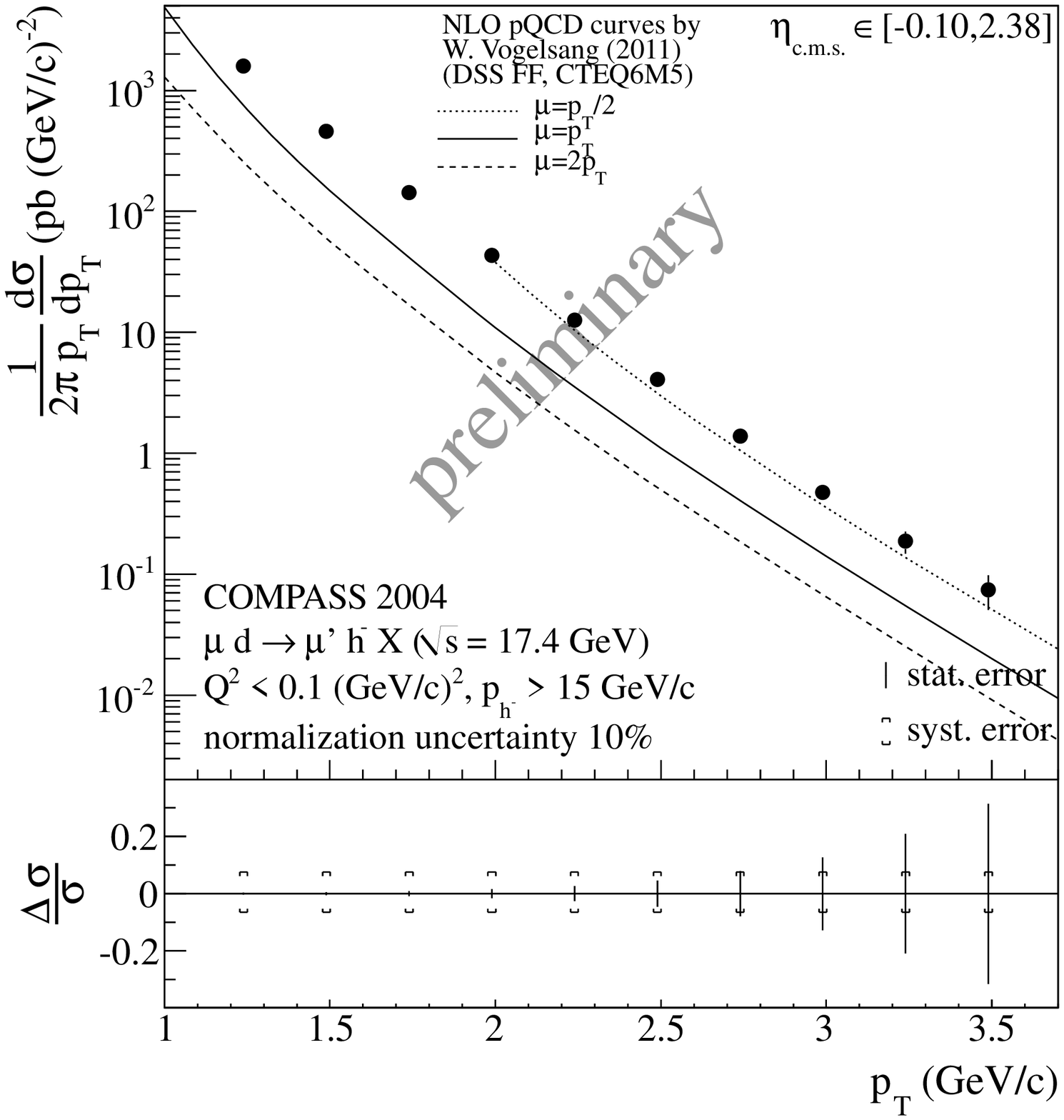}
\vspace{-20pt}\footnotesize
\caption{\label{minus}Cross-section of $h^{-}$ as a function of $p_{T}$ compared to NLO pQCD.}
\normalsize
\vspace{-20pt}

\end{figure}

\begin{figure}
\centering

\includegraphics[width=16pc]{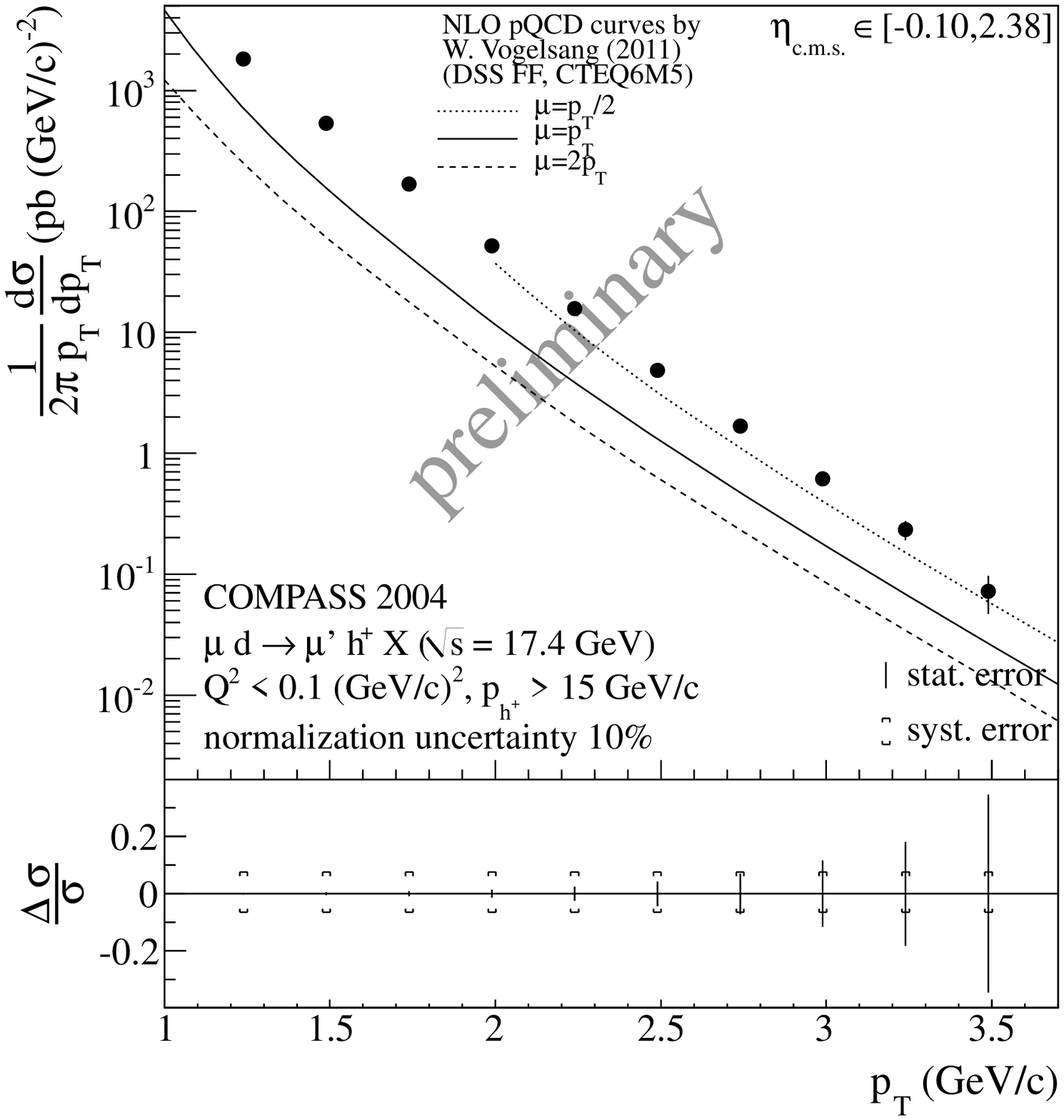}
\vspace{-20pt}
\footnotesize
\caption{\label{plus}Cross-section of $h^{+}$ as a function of $p_{T}$ compared to NLO pQCD.}
\normalsize
\vspace{-20pt}

\end{figure}

\begin{figure}\centering
\includegraphics[width=32pc]{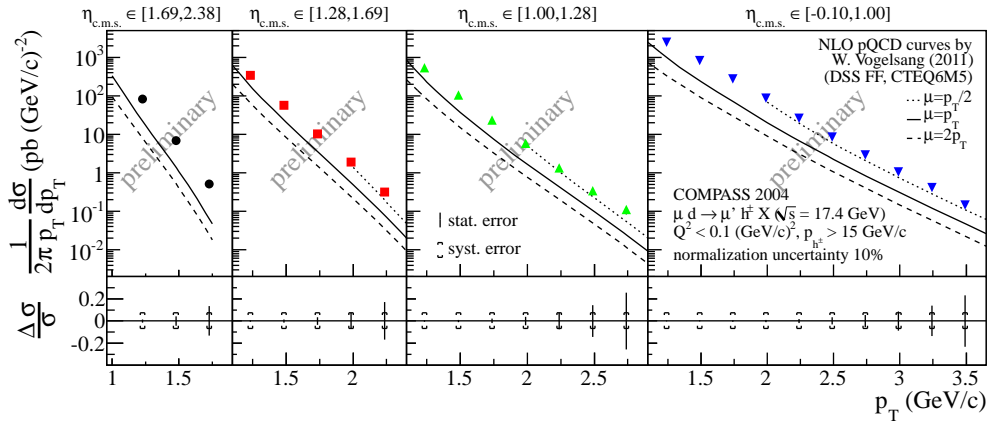}
\caption{\label{eta}Cross-sections of $h^{\pm}$ as a function of $p_{T}$ compared to NLO pQCD in different $\eta$ regions. Leftmost plot shows the most forward accessible $\eta$ in the measurement while the rightmost figure shows the most central $\eta$ value measured.}
\end{figure}

\begin{figure}\centering
\includegraphics[width=16pc]{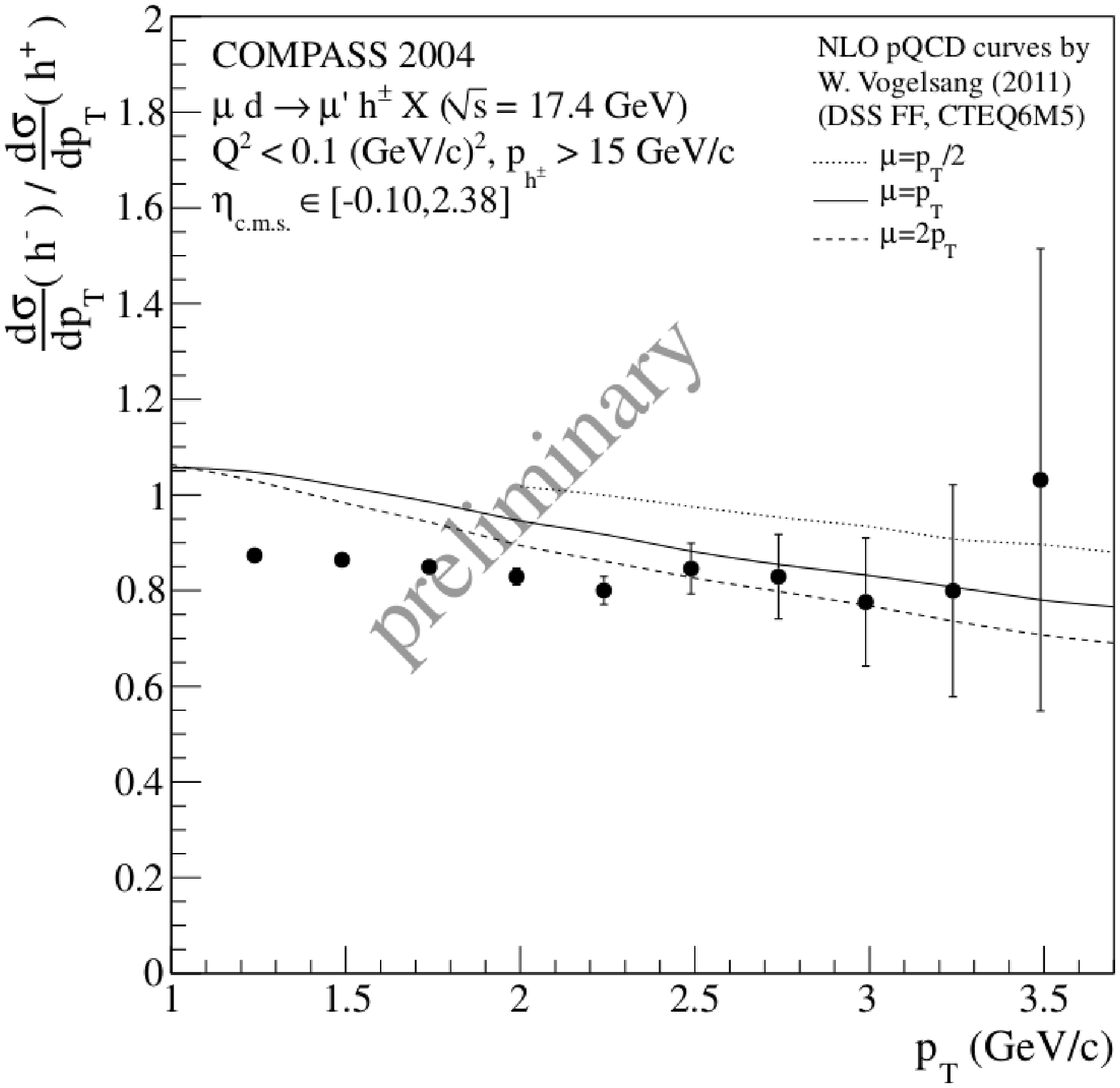}
\footnotesize
\caption{\label{ratio}Ratio of  $h^{-}$ to $h^{+}$ cross-sections compared to NLO pQCD.}
\normalsize
\end{figure}

The measured differential cross-sections for negative and positive hadrons integrated over the $\eta$ range of $-0.10$ to $2.38$ can be found in Fig.~\ref{minus} and~\ref{plus}. The measurements are compared to NLO calculations in which three input scales are used. The measurements show that the data is better described by the input scale of $\mu=2p_{T}$, however given the calculations uncertainty we conclude that the data is over all well described in the integrated $\eta$ range. The cross-sections separated in finer bins of rapidity are shown in Fig.~\ref{eta}. More forward $\eta$ regions appear to have a steeper slope than the more central $\eta$ bins. The more central bins appear to be better described by the theoretical calculations. The ratio of $h^{-}$ to $h^{+}$ cross-sections  is shown in Fig.~\ref{ratio}. The measured ratio shows a $p_{T}$ independent flat distribution  within the measurements errors. This flat trend is in contrast with the theoretical calculations which show a decreasing trend as a function of $p_{T}$. The measurement shows that over all there are more $h^{+}$ in the measurement.



\section{Summary and conclusions}
COMPASS has measured the differential cross sections of charged hadrons integrated over the $\eta$ range of $-0.10$ to $2.38$ and also in finer rapidity bins. The measurements are consistent with the NLO pQCD calculations and within the theory's uncertainties. The ratio of $h^{-}$ to $h^{+}$ show an excess of $h^{+}$ which is independent of $p_{T}$. This ratio indicates a discrepancy on the theory  which predicts more $h^{-}$ at low $p_{T}$ ($<~1.5GeV/c$.) This work is under paper preparation and is partly funded by a personal grant by the American National Science Foundation in collaboration with the French Atomic and Alternative Energies Commission. I thank Werner Vogelsang and Andrei Afanesev for their theoretical calculations which has made the QED and pQCD comparisons to the experimental measurements possible.

\end{document}